\documentclass{svproc}

\pdfoutput=1
\usepackage{url}
\usepackage{cite}
\usepackage{amsmath}
\usepackage{float}
\usepackage{amssymb,amsfonts}
\usepackage{algorithmic}
\usepackage{comment}
\usepackage{lipsum}
\usepackage{graphicx}
\usepackage{textcomp}
\usepackage{siunitx}
\usepackage{xcolor, soul}
\usepackage{hyperref}
\usepackage{cleveref}

\def\alphamu{$\alpha$-$\mu$ }
\def\kappamu{$\kappa$-$\mu$ }
\DeclareSIUnit{\bps}{bps}

\begin{document}
\mainmatter              
\title{Performance Analysis of RIS-Aided NOMA Networks in \alphamu \& \kappamu Generalized Fading Channel}
\titlerunning{Generalized Fading Channels in RIS-Aided NOMA Systems}  
%
\author{Aaditya Prakash Kattekola \and Sanjana Dontha\and
Anuradha Sundru}
\authorrunning{Aaditya Prakash Kattekola et al.} 
%
\tocauthor{Aaditya Prakash Kattekola, Sanjana Dontha, Anuradha Sundru}
\institute{ Department of Electronics \& Communication Engineering\\
National Institute of Technology, Warangal
}
\maketitle              
\begin{abstract}
For forthcoming 5G networks, Non-Orthogonal Multiple Access (NOMA) is a very promising technique. And in today's world, Line of Sight communication is becoming increasingly harder to achieve. Hence, technologies like Reconfigurable Intelligent Surfaces(RIS) emerge. RIS-aided NOMA network is a widely researched implementation of RIS. The environment where these networks are employed are non-homogeneous \& non-linear in nature. The effectiveness of these systems must thus be evaluated using generalized fading channels. In this paper, the performance of a RIS-aided NOMA is compared with conventional NOMA.  in \alphamu \& \kappamu channels. This paper also shows that the well-known fading distribution are special cases of these generalized fading channels, both analytically and through simulation.
\keywords{Reconfigurable Intelligent Surfaces, Non-Orthogonal Multiple Access, Generalized Fading Channels, 5G \& Beyond 5G Networks}
\end{abstract}

\section{Introduction}
\label{ch:intro}
The transmission of speech and data through a wireless network is done without the need of cables or wires. Data is sent through electromagnetic signals that are broadcast from sending facilities to intermediate and end-user devices in place of a physical link.\cite{Moozakis2023-av}. The fifth-generation (5G) technological standard for broadband cellular networks, which cellular phone providers started rolling out globally in 2019, is the anticipated replacement for the 4G networks that connect the majority of modern smart phones.\cite{Mtec2020-yg} For 5G cellular networks, NOMA was suggested as a potential radio access method. Real-time power allocation and consecutive interference cancellation methods for NOMA must be implemented with great computing efficiency in cellular networks.\cite{Kizilirmak2016-fv}. RIS is a new wireless communication technology that uses a flat surface made up of numerous small, programmable reflecting elements to manipulate radio waves in order to improve wireless communication performance. Each element of the RIS can be controlled individually to change the phase of the reflected signal, which enables the surface to reflect, refract, or diffract the incoming signal in a specific direction. Additionally, RIS-assisted  NOMA is viewed as a potential new technology that can reconfigure the environment for wireless transmission via software-controlled reflection to meet the requirements of 5G and Beyond-5G communication.\cite{Ghous2022-mo}.  \\ \\
In wireless communication, the  transmitted signal undergoes large scale \& small scale effects such as reflection, diffraction etc. Hence, the receiver receives multiple components of the signal, each having experienced changes in amplitude, phase and several other properties\cite{Kumbhani2017-pd}. Based on the type of communication, there exist well-known statistical models of these received signal. Rayleigh fading channel is used when the communication is predominantly Non Line-Of-Sight (NLOS) \cite{Sklar1997-ri}. On the other hand, the dominant signal is a Line-Of-Sight (LOS) signal, Rician fading model is used \cite{Tepedelenlioglu2003-uw}. Other channels include Nakagami-m, Hoyt and Weibull. Nakagami-m in particular is widely used due to its generalized nature. When the Nakagami parameter 'm' is set to $1$ \& $0.5$, Rayleigh and One-sided Gaussian are obtained respectively. Yet there exists situations in channel modelling where none of the known channels seem adequate. \cite{Yacoub2007-gj} \\ \\
In the context of RIS-aided NOMA (non-orthogonal multiple access) systems the performance heavily depends on the location and configuration of the RISs. Therefore, the fading model should consider the impact of the RISs on the channel. However, the conventional fading models such as Rayleigh, Rician, or Nakagami-m do not explicitly capture the effect of RISs on the channel \cite{Xie2022-gn}. NOMA networks use superposition coding and successive interference cancellation (SIC) to separate the signals from different users\cite{Kizilirmak2016-fv}. This requires a non-linear signal processing operation that can only be applied under certain conditions. However, these fading channels do not satisfy these conditions and hence may not be suitable for RIS-aided NOMA systems. Secondly, the existing RIS-Assisted channel models lack generality . RISs can significantly modify the channel between the transmitter and the receiver by reflecting the signal in a specific direction. This can result in a non-line-of-sight (NLOS) channel that is different from the conventional LOS channel model. As a result, the conventional fading models such as Rayleigh, Rician, or Nakagami-m may not accurately capture the characteristics of the RIS-aided channel\cite{Huang2022-no}. \\ \\
This paper aims to analyse the performance of RIS-assisted NOMA system under two generalized fading channels. Namely, the \alphamu \& \kappamu fading channels. Particularly, the outage probability of the system is simulated. In \Cref{ch:model} a brief description of the system model is given. \Cref{ch:method} \& \Cref{ch:results} goes through \alphamu \& \kappamu fading channels and their numerical simulation.
\section{System Model}
\label{ch:model}
\vspace{-7mm}
\begin{figure}[H]
    \centering
    \includegraphics[width=0.45\textwidth]{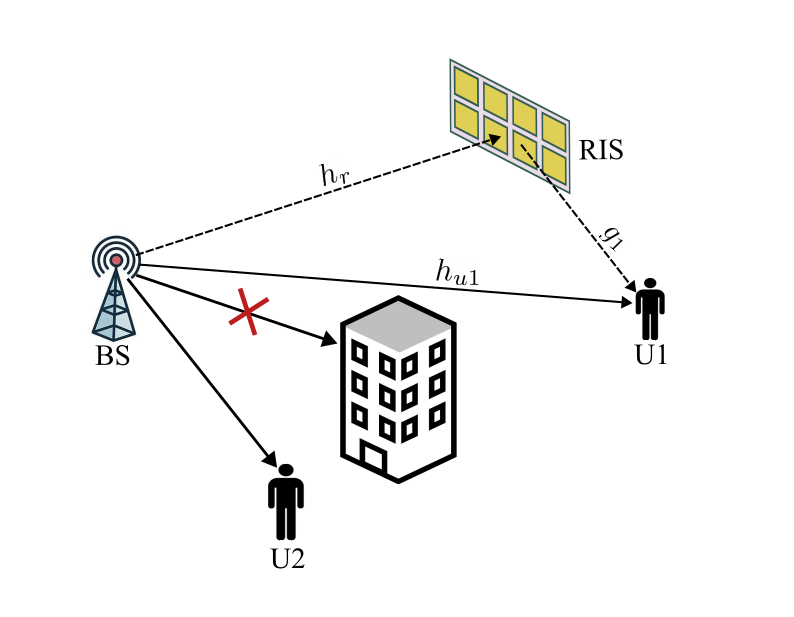}
    \caption{RIS-aided NOMA System}
    \label{fig:ris aided noma}
\end{figure}
\Cref{fig:ris aided noma} shows the system model. 
It is believed that the nearby user (U2) is so close to the base station (BS) that U2 receives all of its signals from the BS. It is possible to represent the channel between the BS and U2 as Rayleigh fading. The Nakagami-m fading channel may be used to represent the BS-RIS connection, which can be either LOS or NLOS. As U1 gets both direct BS signals and reflected RIS signals, it will be our primary user of interest. The Nakagami-m fading channel may be used to represent the RIS-U1 connection, which can also be either a LOS or NLOS link. \cite{Hou2020-mt}.
The signal received by U1 is: 
\begin{equation}
h_{1} = 
h_{u1} d_1^{\frac{-\eta_n}{2}} + g_1\Phi h_r^H  d_{ris}^{\frac{-\eta_l}{2}}  d_{r,1}^{\frac{-\eta_n}{2}} 
\label{eq:signal}    
\end{equation}
where $\Phi$ is the phase shift matrix of the RIS which models the effective phase shift applied by all intelligent surfaces. Mathematically,  $\Phi$ is a diagonal matrix with each element is of form $\beta_ie^{j\phi_i}$. $h_{u1}$ denotes the signal received by U1 from BS,  $d_{1}$ denotes the distance between BS and U1,
$h_{r}$ denotes the signal received by RIS from BS, $d_{ris}$ denotes the distance between BS and RIS,
and $d_{r,1}$ denotes the distance between RIS and U1. $\eta_n\ \text{and}\ \eta_l$ denote the path loss factor for the BS to U1 and BS-RIS to U1 path, respectively.
\section{Generalized Fading Channels}
\label{ch:method}
\subsection{$\kappa$-$\mu$ Channel}
\kappamu fading channel has become popular in analytical research of wireless communication scenarios.
\kappamu distribution is suitable to model LOS environments just like the Nakagami. In non-homogeneous environments where reflecting \& scattering obstacles of different dimensions are present, \kappamu statistics can be used to model the scenario \cite{Kumbhani2017-pd}. 
\subsubsection{Probability Density Function (PDF) of \kappamu Channel}
{\small\begin{equation}
\begin{aligned}
p_X(x) = \\   
\frac{\mu(1+\kappa)^{\frac{\mu+1}{2}}}{\kappa^{\frac{\mu-1}{2}}e^{\mu\kappa}\bar{X}^{\frac{\mu+1}{2}}}
& x^{\frac{\mu-1}{2}}
e^{-\frac{\mu(1+\kappa)x}{\bar{X}}}
I_{\mu-1}\left(2\mu\sqrt{\frac{\kappa(1+\kappa)x}{\bar{X}}}\right)
\end{aligned}
\label{eq:PDF of K Mu}
\end{equation}
}
\subsubsection{Cumulative Density Function (CDF) of \kappamu Channel}
\begin{equation}    
P_x(y) = 
1 - 
Q_u
\left(
    \sqrt{2\kappa\mu},
    \sqrt{\frac{2\mu(1+\kappa)y}{\bar{X}}}
\right)
\label{eq:CDF of K Mu}
\end{equation}
\subsubsection{Moment Generating Function (MGF) of \kappamu Channel}
\begin{equation} 
M_X(s) = 
\left(
    \frac{\mu(1+\kappa)}{\mu(1+\kappa)+s\bar{X}}
\right) ^\mu
e^{\frac{\mu^2\kappa(1+\kappa)}{\mu(1+\kappa)+s\bar{X}}-\kappa\mu}
\label{eq:MGF of K Mu}
\end{equation}
\begin{figure}[H]
    \centering
    \includegraphics[width=0.45\textwidth]{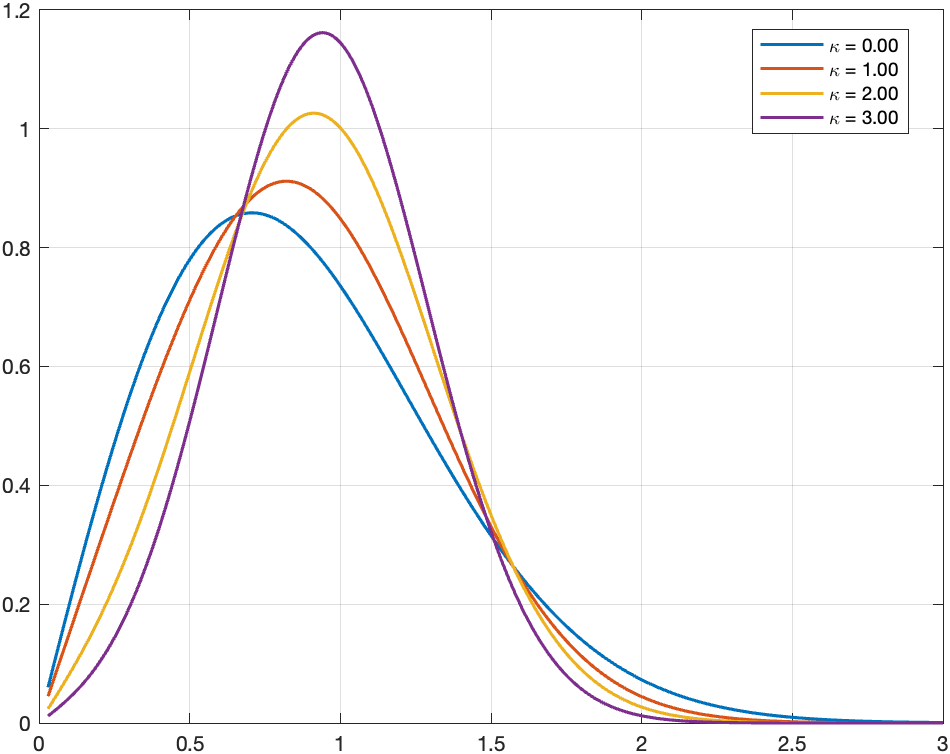}
    \includegraphics[width=0.45\textwidth]{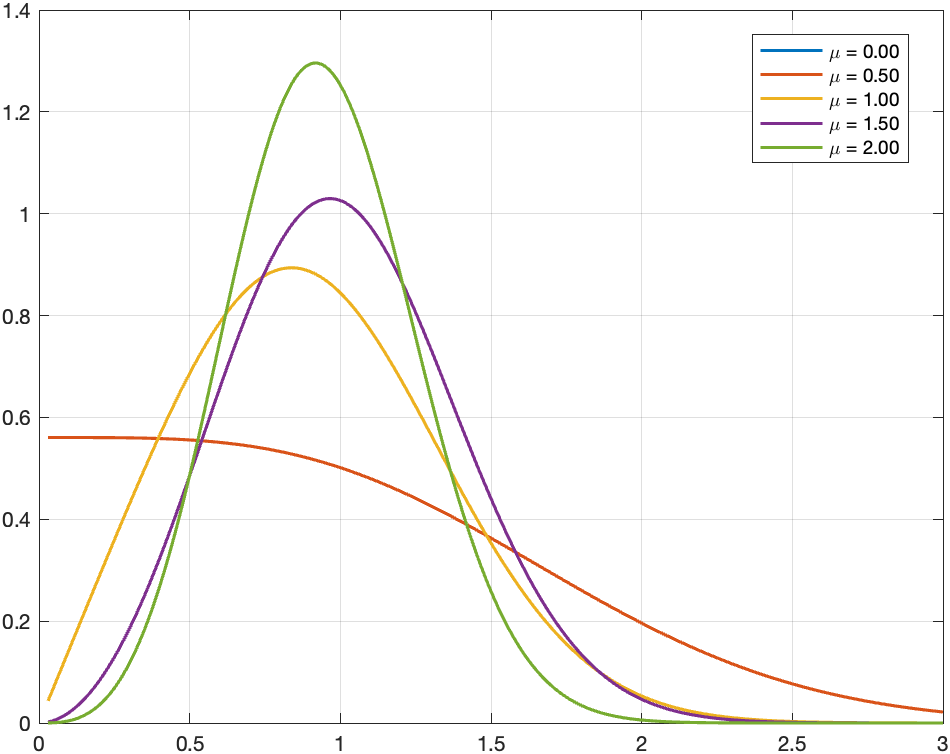}
    \caption{The PDF of \kappamu for various $\kappa$ \& $\mu$}
    \label{fig:kappa mu varying}
\end{figure}
\subsection{$\alpha$-$\mu$ Channel}
\alphamu channel is a non-linear Nakagami-m channel. In non-homogeneous environments that introduce non-linearity in the signal, \alphamu fading channel can be employed \cite{Kumbhani2017-pd}. 
\subsubsection{Probability Density Function (PDF) of \alphamu Channel}
\begin{equation}
p_X(x) = \frac{\alpha \mu^\mu x^{\alpha\mu-1}}{\Gamma(\mu)\Omega_\alpha^{\alpha\mu}}e^{-\mu\left(\frac{x}{\Omega_\alpha}\right)^\alpha}, \mu, \alpha > 0, x \ge 0
\label{eq:PDF of A Mu}
\end{equation}
\subsubsection{Cumulative Density Function (CDF)of \alphamu Channel:}
\begin{equation}
P_x(y) = 
\frac{\gamma\left(\mu,\mu\left(\frac{y}{\Bar{x}}\right)^\frac{\alpha}{2}\right)}{\Gamma(\mu)}
\label{eq:CDF of A Mu}
\end{equation}
\subsubsection{Moment Generating Function (MGF) of \alphamu Channel}
{\footnotesize \begin{equation}
  \begin{aligned}
     M_x(s) = \\ 
    \frac{\alpha\mu^\mu}{2\bar{x}^\frac{\alpha\mu}{2}}
\frac{\sqrt{k}l^{\frac{\alpha\mu-1}{2}}}{(2\pi)^{\frac{l+k-2}{2}}s^\frac{\alpha\mu}{2}}
       & G_{l,k}^{k,l}
\left(
\left(\frac{\mu}{\bar{x}^\frac{\alpha}{2}}\right)^k \frac{l^l}{s^lk^k}
,\begin{array}{c}
     P(l,1-\frac{\alpha\mu}{2}) \\   P(k,0) 
\end{array}\right) \\
  \end{aligned}
\label{eq:MGF of A Mu}
\end{equation}
}
\begin{figure}[H]
    \centering
    \includegraphics[width=0.45\textwidth]{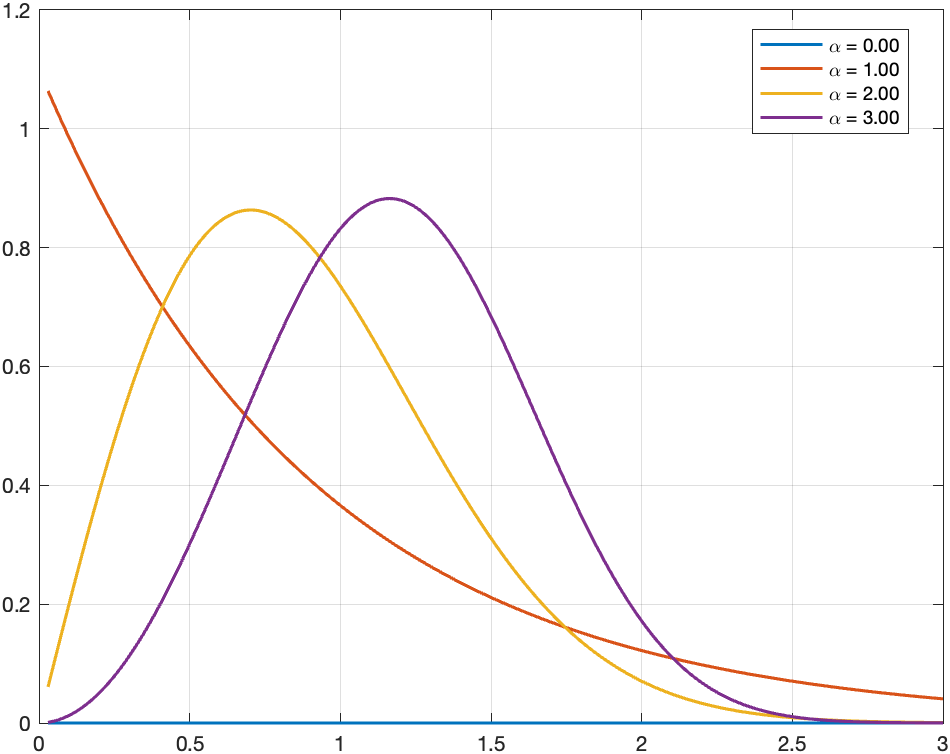}
    \includegraphics[width=0.45\textwidth]{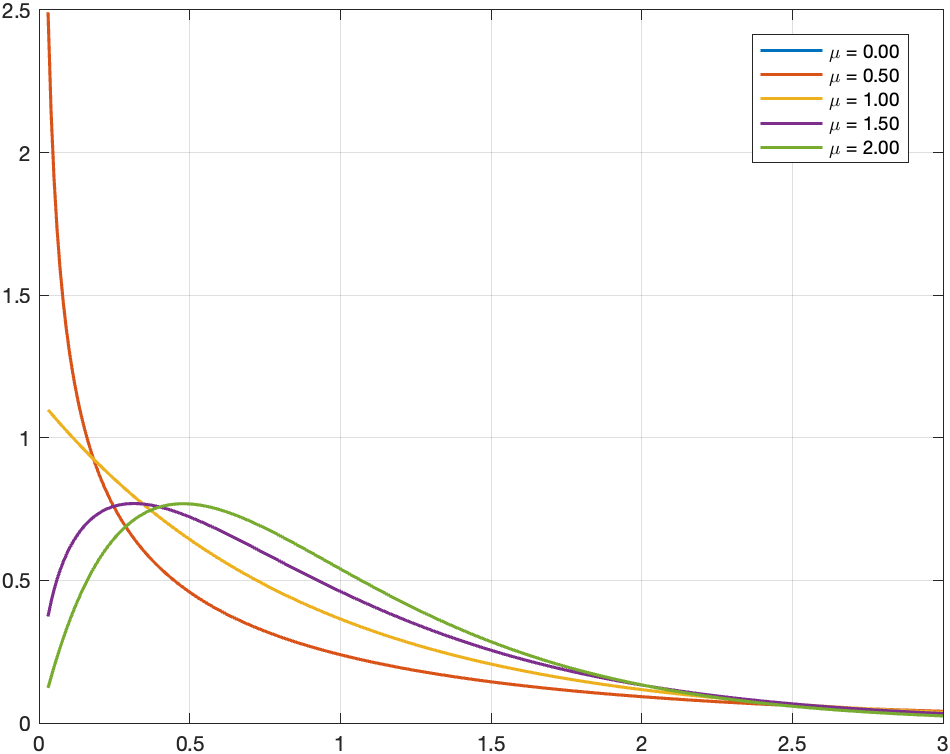}
    \caption{The PDF of \alphamu for various $\alpha$ \& $\mu$}
    \label{fig:alpha mu varying}
\end{figure}
\section{Simulation \& Results}
\label{ch:results}
\subsection{Simulation Setup}
\label{subch:simulation setup}
The downlink's carrier frequency and bandwidth are set to BW = 100 Hz.\si{\mega\hertz} and 1\si{\giga\hertz}, respectively. The AWGN's power is set to $\sigma^2 = - 174 + 10 \cdot \text{log}_{10}(BW)$ \si{\decibel}m. At the reference distance, the power attenuation  is set to -30 \si{\decibel}, and 1 meter is chosen as the reference distance value.Be aware that the Nakagami fading parameter identifies the LOS and NLOS connections, where $ m_{g1} = 1$ and $ m_{g1} > 1$ are for NLOS and for LOS links, respectively. The target rates are $R1$ = 1.5 and $R2$ = 1 \si{\bps\per\hertz}. The paired NOMA users' power allocation factors are set to $\alpha^2_2 = 0.25\ \text{and}\ \alpha^2_1 = 0.75$. The direct BS-user connections and the reflected BS-RIS-route user's loss exponents are both set to $\eta_l = 2.2\ \text{and}\ \eta_n = 3.5$, respectively. The signal alignment method was used to determine the phase shift matrix.\cite{Hou2020-mt}.
\subsection{Results}
\begin{figure}[H]
    \centering
    \includegraphics[width=0.38\textwidth]{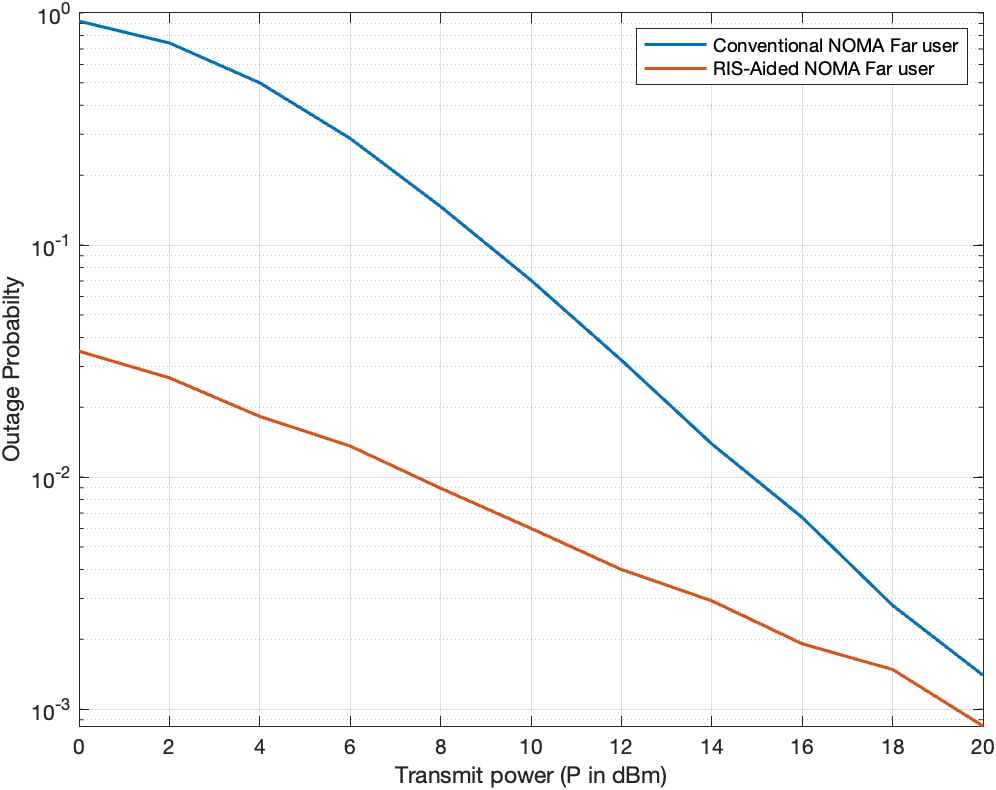}
    \caption{Outage Probability in a Nakagami-m Fading}
    \label{fig:OP ground truth}
\end{figure}

\begin{figure}[H]
    \centering
    \begin{minipage}{0.47\textwidth}
      \centering
    \includegraphics[width =1.01 \textwidth]{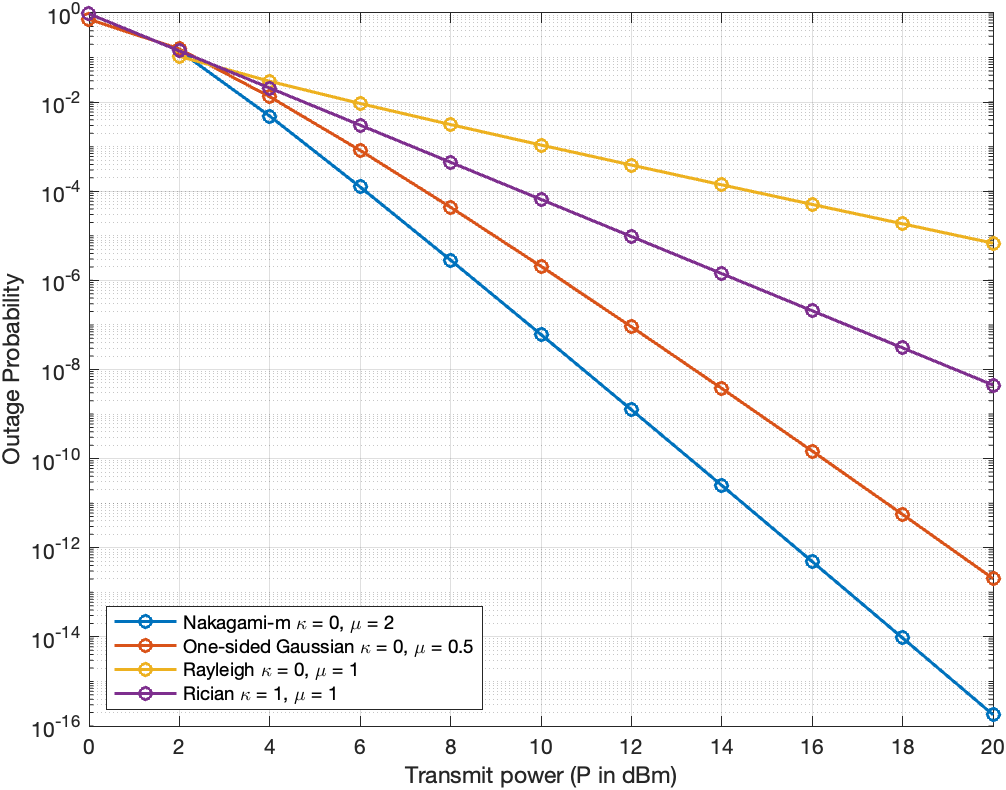}
    \caption{Outage Probability of\\ Generalized $\kappa$-$\mu$ channel}
    \label{fig:OP k mu}
    \end{minipage}%
\hspace{1mm}
    \begin{minipage}{0.47\textwidth}
        \centering
    \includegraphics[width = 1.01\textwidth]{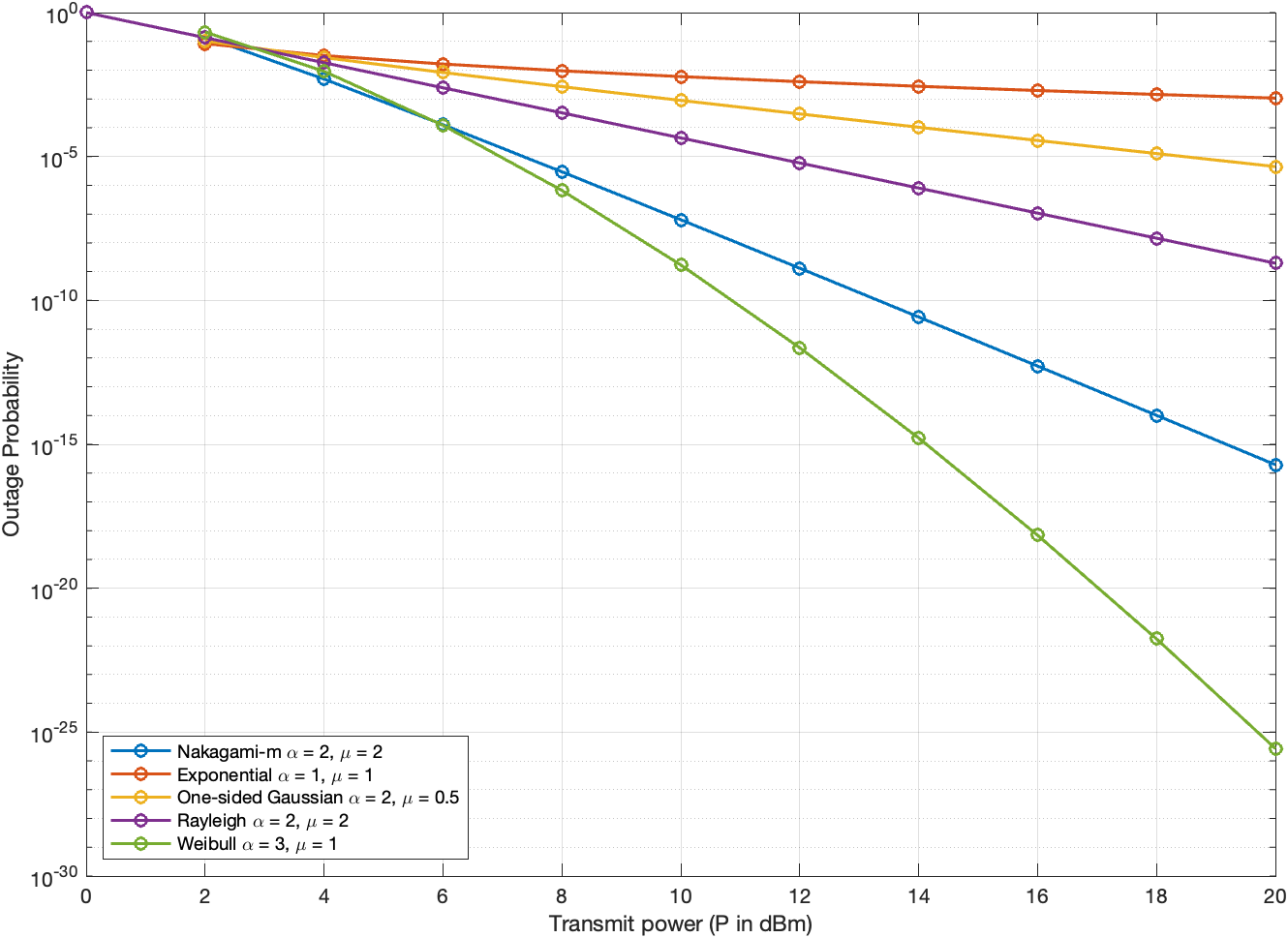}
    \caption{Outage Probability of\\ Generalized $\alpha$-$\mu$ channel}
    \label{fig:OP A mu}
    \end{minipage}
\end{figure}

\begin{figure}[H]
    \centering
    \includegraphics[width=0.6\textwidth]{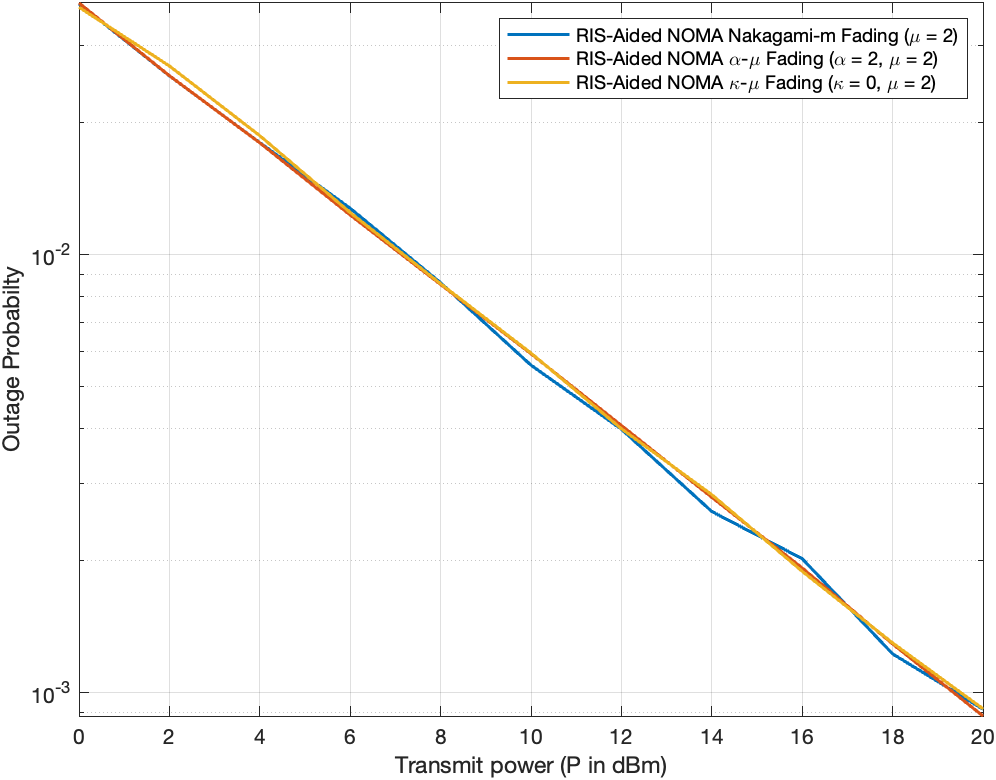}
    \caption{Nakagami-m Channel using Generalized \alphamu \& \kappamu Channels }
    \label{fig:a mu k mu as nakagami}
\end{figure}

Outage probability (OP) is defined as the proboability that the SNR at the receiver falls below a certain threshold \cite{ZHANG201719}.
In this section, OP of a conventional NOMA and a RIS-assisted NOMA network are compared for performance analysis. This comparison is followed by the showing that the well-known channels are shown as special cases of \alphamu \& \kappamu models. To prove this point further, the OP of a Nakagami-m fading channel is compared with \alphamu \& \kappamu versions of the Nakagami. 
\Cref{fig:OP ground truth} shows that the RIS-assisted NOMA outperforms conventional NOMA. This result is rational as U1 receives a stronger signal and better directed signal in the presence of the RIS. It should be noted that, the RIS should be tuned appropriately.
The generalized fading channels, as mentioned earlier, can be use to generate many well-known fading channels.
In the case of \alphamu, we can generate Nakagami-m, Rayleigh, and Weibull to name a few. In \Cref{fig:OP A mu} \& \Cref{tab:A Mu parameters} the specific $\alpha$ and $\mu$ values for these distributions is shown. 
\begin{table}[H]
    \centering
    \begin{minipage}{0.47\textwidth}
        \centering
   \begin{tabular}{|c|c|c|}
    \hline
        \textbf{Distribution} & \textbf{Alpha($\alpha$)} & \textbf{Mu($\mu$)}\\
         \hline
         Exponential & 1 & 1\\
         \hline
         Weibull & 3 & 1\\
         \hline
         Nakagami-m & 2 & 2\\
         \hline
         Rayleigh & 2 & 1\\
         \hline
         One-sided Gaussian & 2 & 0.5\\
         \hline
    \end{tabular}
    \vspace{2.5mm}
    \caption{\alphamu values for well-known distributions}
    \label{tab:A Mu parameters}
    \end{minipage}%
\hspace{1mm}
    \begin{minipage}{0.47\textwidth}
    \centering
   \begin{tabular}{|c|c|c|}
    \hline
        \textbf{Distribution} & \textbf{Kappa($\kappa$)} & \textbf{Mu($\mu$)}\\
         \hline
         Rice & 1 & 1\\
         \hline
         Nakagami-m & 0 & 2\\
         \hline
         Rayleigh & 0 & 1\\
         \hline
         One-sided Gaussian & 0 & 0.5\\
         \hline
    \end{tabular}
    \vspace{2.5mm}
    \caption{\kappamu values for well-known distributions}
    \label{tab:K Mu parameters}
    \end{minipage}
\end{table}
Similar to \alphamu, \kappamu distribution can model Nakagami-m, Rayleigh \& Rice distributions. The values for $\kappa$ and $\mu$ to replicate these fading channels is given in \Cref{fig:OP k mu} \& \Cref{tab:K Mu parameters}.
Finally, the OP for the physical model described in \cref{subch:simulation setup} is simulated using Nakagami-m channel and also \alphamu \& \kappamu generated Nakagami channel.
\Cref{fig:a mu k mu as nakagami} clearly shows that when $alpha = 2\ \text{and}\ \kappa = 0$ for \kappamu and \alphamu channels, respectively, they replicate the OP of a Nakagami-m distribution almost perfectly
\section{Conclusion}
\label{ch:conclusion}
The RIS-Aided NOMA system is modelled and its performance (in-terms of outage probabilty) was compared and it was shown that RIS-aided NOMA performance better for U1. The utility of \alphamu \& \kappamu channels in generating well-known fading channels was also shown. It was also demonstrated that these channels can replicate a Nakagami-m channel almost perfectly. 
%
%
\bibliographystyle{splncs_srt}
\bibliography{ms}
\end{document}